\documentclass[12pt]{article}

\usepackage{amsmath, euscript, amssymb, amsfonts, latexsym}
\usepackage{mathrsfs}

\newcommand{\mS}{{\mathscr{S}}}
\newcommand{\mW}{{\mathscr{W}}}

\newcommand{\oR}{{\mathbb R}}
\newcommand{\oV}{{\mathbb V}}
\newcommand{\oC}{{\mathbb C}}
\newcommand{\oZ}{{\mathbb Z}}

\newcommand{\eqdef}{\stackrel{\mathrm{def}}{=}}

\begin{document}

\baselineskip=20pt

\hfill  FIAN/TD/16-07

\hfill J.Phys.A: Math.Theor. 40 (2007) 14593-14604 \vspace{2cm}

\begin{center}

{\Large\bf Noncommutativity and $\theta$-locality}

\vspace{1cm}

{\bf M.~A.~Soloviev}\footnote{E-mail: soloviev@lpi.ru}

\vspace{0.5cm}

 \centerline{\sl P.~N.~Lebedev Physical Institute}
 \centerline{\sl Russian Academy of Sciences}
 \centerline{\sl  Leninsky Prospect 53, Moscow 119991, Russia}

\vskip 3em

\end{center}

\begin{abstract}
In this paper, we introduce the condition of $\theta$-locality
which can be used as a substitute for microcausality in quantum
field theory on noncommutative spacetime. This condition is
closely related to the asymptotic commutativity which was
previously used in nonlocal QFT. Heuristically, it means that the
commutators of observables behave at large spacelike separation
like $\exp(-|x-y|^2/\theta)$, where $\theta$ is the
noncommutativity parameter. The rigorous formulation given in the
paper implies averaging fields with suitable test functions. We
define a test function space  which most closely corresponds to
the Moyal $\star$-product and prove that this space is a
topological algebra under the star product. As an example, we
consider the simplest normal ordered monomial $:\phi\star\phi:$
and show that it obeys the $\theta$-locality condition.
\end{abstract}

\vskip 2em

\section{\large Introduction}
In this paper, we discuss the problem of formulating the causality
principle in quantum field theory on noncommutative spacetime. A
 noncommutative spacetime of $d$  dimensions  is defined by
replacing the coordinates  $x^\mu$ of $\oR^d$ by Hermitian
operators $\hat x^\mu$ satisfying the commutation relations
\begin{equation}
[\hat x^\mu, \hat x^\nu]=i\theta^{\mu\nu},
 \label{1.1*}
\end{equation}
where $\theta^{\mu\nu}$ is a real antisymmetric  $d\times d$
matrix, which will henceforth be assumed constant as in most
papers on this subject. The Weyl-Wigner correspondence between
algebras of operators and  algebras of functions enables one to
consider  quantum field theories on noncommutative spacetime as a
form of nonlocal QFT described by an action, in which the ordinary
product of fields is replaced by the Moyal-Weyl-Groenewold star
product
\begin{multline}
(f\star_\theta
g)(x)=f(x)\exp\left(\frac{i}{2}\,\overleftarrow{\partial_\mu}\,
\theta^{\mu\nu}\,\overrightarrow{\partial_\nu}\right)g(x)\\
=f(x)g(x)+\sum_{n=1}^\infty\left(\frac{i}{2}\right)^n\frac{1}{n!}\,
\theta^{\mu_1\nu_1}\dots
\theta^{\mu_n\nu_n}\partial_{\mu_1}\dots
\partial_{\mu_n}f(x)\partial_{\nu_1}\dots\partial_{\nu_n}g(x)
\label{1.2*}
\end{multline}
(see, e.g.,~\cite{Sz} for more details). Recent interest in
noncommutative QFT was  caused mainly by the fact that it occupies
an intermediate position between the usual quantum field theory
and string theory~\cite{SeiW}. At present, considerable study is
being given not only to actual models, but also  to the conceptual
framework of this theory. In particular, in~\cite{Alv, Ch1,FP,FW}
efforts were made to derive a corresponding generalization of the
axiomatic approach~\cite{SW,J,BLOT}. Much attention is being given
to the nonlocal effects inherent in noncommutative QFT. A
comparison of the theories in which the time coordinate is
involved in the noncommutativity and the theories with
$\theta^{0\nu}=0$ shows that these latter are preferable because
they obey unitarity. In~\cite{Alv,LS,CFI}, it was argued, however,
that in the case of space-space noncommutativity the usual causal
structure with the light cone is replaced  by a structure with a
light wedge respecting the unbroken $SO(1,1)\times SO(2)$
symmetry. Since quantum fields are singular by their very nature,
a comprehensive study of the question of causality must include
finding an adequate space of test functions. In the standard
formalism~\cite{SW,J,BLOT},  quantum fields are taken to be
tempered operator-valued distributions, which are defined on  the
Schwartz space $\mS$ consisting of all infinitely differentiable
functions of fast decrease. As noted in~\cite{Alv}, the assumption
of temperedness is open to the  question in noncommutative QFT
because of UV/IR mixing. Moreover, the correlation  functions of
some gauge-invariant operators admit an exponential growth at
energies much larger than the noncommutativity scale~\cite{GHI},
and this is an argument in favour of analytic test functions. The
very structure of the star product~\eqref{1.2*}, which is defined
by an infinite-order differential operator, suggests that analytic
test functions may be used in noncommutative QFT along with or
instead of  Schwartz's $\mS$.

In~\cite{S07}, we argued that the appropriate test function spaces
 must be algebras under the Moyal
$\star$-product and showed that the  spaces $S^\beta_\alpha$ of
Gelfand and Shilov~\cite{GS2} satisfy this condition if and only
if $\alpha\ge\beta$. The space $S^\beta_\alpha$ consists of the
smooth functions that decrease at infinity faster than
exponentially with order $1/\alpha$  and a finite type, and whose
Fourier transforms behave analogously but with order $1/\beta$.
Clearly, all these spaces are contained in the space $\mS$, which
can be thought of as $S^\infty_\infty$. As shown in~\cite{S07},
the series~\eqref{1.2*} is absolutely convergent for any $f,g\in
S^\beta_\alpha$ if and only if $\beta<1/2$. However, the  star
multiplication  has a unique continuous extension to any space
$S^\beta_\alpha$ with $\alpha\ge\beta$. It is natural to use the
spaces with $\beta<1/2$  as an initial functional domain  of
quantum fields on noncommutative spacetime, but this does not rule
out a possible extension to a larger test function space depending
on the model under consideration. Recently, the use of spaces
$S^\beta=S^\beta_\infty$, $\beta<1/2$, was also advocated by
M.~Chaichian {\it et al}~\cite{Ch2}.

If $\beta<1$, then the test functions are entire analytic, and the
notion of support loses its meaning for the generalized functions
that are defined on $S^\beta_\alpha$ or $S^\beta$ and constitute
their dual spaces $S^{\prime\beta}_\alpha$ and $S^{\prime\beta}$.
Nevertheless, some basic theorems of the theory of distributions
can be extended to these generalized functions because they retain
the property of angular localizability~\cite{FS92,S93}. This
 property leads naturally to the condition of asymptotic
commutativity, which was used in nonlocal QFT  instead of local
commutativity and was shown to ensure the existence of
CPT-symmetry and the standard spin-statistics relation for
nonlocal fields~\cite{S99}.  We  already discussed in~\cite{S06}
how some of these proofs with test functions in $S^0$ can be
adapted to noncommutative QFT. Here we intend to argue that
quantum fields living on  noncommutative spacetime indeed satisfy
the asymptotic commutativity condition and to explain the
interrelation between this condition and the fundamental length
scale which is determined by the noncommutativity
parameter~$\theta$. To avoid notational clutter, we will use the
one-index spaces of type $S$, although the two-index spaces
provide a wider distributional framework.

In section~2, we introduce the test function space
$\mathscr{S}^{1/2}$ which most closely corresponds to the Moyal
star product. All spaces $S^\beta$ with $\beta<1/2$ are contained
in this space, but it is smaller than $S^{1/2}$ and may be defined
as a maximal space with the property that the series~\eqref{1.2*}
is absolutely convergent for any pair of its elements. We also
prove that $\mathscr{S}^{1/2}$ is a topological algebra under the
 star product. In section~3, two classes of spaces related to
$S^{\beta}$, $\mathscr{S}^{\beta}$ and associated with cones in
$\oR^d$ are defined, and it is shown that  these spaces  are
algebras under the $\star$-product for $\beta<1/2$ in the former
case and for $\beta\le 1/2$ in the latter case. In section~4, the
exact formulation of the asymptotic commutativity condition is
given and its physical consequences are briefly outlined. In the
same section we introduce the notion of $\theta$-locality. In
section~5, we take,  as a case in point, the normal ordered
$\star$-square $:\phi\star\phi:$ of the free scalar field $\phi$
and show that it obeys the conditions of asymptotic commutativity
and $\theta$-locality. Section~6 contains concluding remarks.

\section{\large Test function spaces adequate to the Moyal  star product}

An advantage of the spaces $S^\beta$ over $\mS$ is  their
invariance under the action of infinite-order differential
operators, the set of which increases with decreasing $\beta$. In
what follows, we consider functions defined on $\oR^d$ and use the
usual multi-index notation:
$$
\partial^\kappa=\displaystyle{\frac{\partial^{|\kappa|}}{\partial
x_1^{\kappa_1}\dots\partial x_d^{\kappa_d}}},\qquad
|\kappa|=\kappa_1+\dots+\kappa_d,\qquad
\kappa^\kappa=\kappa^{\kappa_1}_1\cdot\dots\cdot
\kappa_d^{\kappa_d},
$$
where $\kappa\in \oZ_+^d$.  Let $\beta\ge0$, $B>0$, and $N$ be an
integer. We denote by $S_N^{\beta,B}(\oR^d)$ the Banach space of
infinitely differentiable function with the norm
\begin{equation}
\|f \|_{B,N}=\sup_{x,\kappa}\,(1+|x|)^N\frac{|\partial^\kappa
f(x)|}{B^{|\kappa|}\kappa^{\beta\kappa}}.
 \label{2.1*}
\end{equation}
 We also
write $S_N^{\beta,B}$ for this space when  this cannot lead to
confusion. Let us consider the operator
\begin{equation}
\sum_{\lambda\in \oZ_+^d} c_\lambda \partial^\lambda
 \label{2.2*}
\end{equation}
assuming that $\sum_\lambda c_\lambda z^\lambda$  has less than
exponential growth of order $\le 1/\beta$ and type $b$. In
treatise~\cite{GS2}, it was shown that under the condition
$b<\beta/(e^2B^{1/\beta})$ the operator~\eqref{2.2*} maps the
space $S_N^{\beta,B}$ to $S_N^{\beta,B'}$, where $B'=e^\beta B$.
This result can  be improved by using the inequality
$(k+l)^{k+l}\le 2^{k+l}k^kl^l$. The assumption of order of growth,
together with the Cauchy inequality, implies that $|c_\lambda|\le
C \prod_{j=1}^d r_j^{-\lambda_j}e^{b\,r_j^{1/\beta}}$ for any
$r_j>0$. Locating the minimum with respect to $r_j$, we obtain
\begin{equation}
|c_\lambda|\le
C\left(\frac{be}{\beta}\right)^{\beta|\lambda|}\frac{1}
{\lambda^{\beta\lambda}}.
\label{2.3*}
\end{equation}
If $f\in S_N^{\beta,B}$, then we have
\begin{multline}
(1+|x|)^N\left|\partial^\kappa\sum_\lambda
c_\lambda\partial^\lambda f(x)\right|\le \|f \|_{B,N}\sum_\lambda
c_\lambda B^{|\kappa+\lambda|}(\kappa+\lambda)^{\beta(\kappa+\lambda)} \\
\leq \|f
\|_{B,N}2^{\beta|\kappa|}B^{|\kappa|}\kappa^{\beta\kappa}\sum_\lambda
c_\lambda 2^{\beta|\lambda|}B^{|\lambda|}\lambda^{\beta\lambda}.
 \label{2.4*}
\end{multline}
Suppose that
\begin{equation}
b<\frac{\beta}{2eB^{1/\beta}}.
 \label{2.5*}
\end{equation}
Then the last series in~\eqref{2.4*}  converges by virtue of the
inequality~\eqref{2.3*}. Taking  $B'\ge 2^\beta B$, we obtain
$\|\sum_\lambda c_\lambda\partial^\lambda f\|_{B',N}\le C'\|f
\|_{B,N}$ and conclude that the operator~\eqref{2.2*} maps
$S_N^{\beta,B}$ to $S_N^{\beta,B'}$ continuously.

Now we apply this consideration to the operator
\begin{equation}
\exp\left(\frac{i}{2}\,\theta^{\mu\nu}\frac{\partial}{\partial
x_1^\mu} \frac{\partial}{\partial
x_2^\nu}\right)=\sum_{n=0}^\infty\left(\frac{i}{2}\right)^n\frac{1}{n!}\,
\theta^{\mu_1\nu_1}\dots\theta^{\mu_n\nu_n}\frac{\partial}{\partial
x_1^{\mu_1}}\dots\frac{\partial}{\partial
x_1^{\mu_n}}\frac{\partial}{\partial
x_2^{\nu_1}}\dots\frac{\partial}{\partial x_2^{\nu_n}}.
 \label{2.6*}
\end{equation}
Clearly, the order of the entire function
$\exp((i/2)\theta^{\mu\nu}z_{1\mu} z_{2\nu})$ is equal to $2$ and
the type is less than or equal to $|\theta|/4$, where
$$
|\theta|= \sum_{\mu<\nu}|\theta^{\mu\nu}|.
$$
Hence we have the following theorem.

\medskip
 {\bf Theorem 1.}
{\it Let $B<1/\sqrt{e|\theta|}$. Then the operator~\eqref{2.6*}
maps the space $S_N^{1/2, B}(\oR^{2d})$ continuously into the
space $S_N^{1/2,B'}(\oR^{2d})$, where $B'= B\sqrt{2}$. The series
obtained by applying this operator to a function $f\in S_N^{1/2,
B}(\oR^{2d})$ is absolutely convergent in the norm
$\|\cdot\|_{B',N}$.}

\medskip
We define the countably-normed  spaces $\mS^\beta$ by
\begin{equation}
\mathscr{S}^\beta=\bigcap_{N,B}S^{\beta,B}_N.
 \label{2.7*}
\end{equation}
A sequence $f_n$   converges to $f\in\mathscr{S}^\beta$ if
$\|f_n-f\|_{B,N}\to 0$ for every $B>0$ and for every~$N$. The
foregoing  leads directly to the following result.

\medskip
 {\bf Theorem 2.} {\it The operator~\eqref{2.6*} maps
the space $\mathscr{S}^{1/2}(\oR^{2d})$ to itself continuously.
Hence it is  well defined and continuous on its dual space
$\mathscr{S}^{\prime 1/2}(\oR^{2d})$. The series obtained by
applying this operator to $f\in \mathscr{S}^{1/2}(\oR^{2d})$ is
absolutely convergent in each of the norms of
$\mathscr{S}^{1/2}(\oR^{2d})$.}

\medskip
Below is given a description  in terms of the Fourier transform,
which shows that the operator~\eqref{2.6*} is  bijective on
$\mathscr{S}^{1/2}$ and so it  is a linear topological isomorphism
of $\mathscr{S}^{1/2}$ as well as of $\mathscr{S}^{\prime 1/2}$.
Analogous statements certainly hold for any $\mathscr{S}^\beta$
with $\beta\le 1/2$, but $\mathscr{S}^{1/2}$ is the largest of
these spaces and most closely corresponds to the
operator~\eqref{2.6*} and hence to the Moyal product~\eqref{1.2*}.
In what follows, we use the notation
$$
\partial_{x_1}\theta\,\partial_{x_2}= \theta^{\mu\nu}
\frac{\partial}{\partial x_1^\mu} \frac{\partial}{\partial
x_2^\nu}.
$$
The map $\mathscr{S}^\beta(\oR^d)\times
\mathscr{S}^\beta(\oR^d)\to\mathscr{S}^\beta(\oR^d)$ that takes
each pair $(f,g)$ to the function $f\star g$ can be considered as
the  composite map
\begin{equation}
\mathscr{S}^\beta(\oR^d)\times
\mathscr{S}^\beta(\oR^d)\stackrel{\otimes}{\longrightarrow}\mathscr{S}^\beta(\oR^{2d})
\stackrel{e^{(i/2)\partial_{x_1}\theta\,\partial_{x_2}}}
{\longrightarrow}\mathscr{S}^\beta(\oR^{2d})\stackrel{\widehat
{\mathsf m}}{\longrightarrow}\mathscr{S}^\beta(\oR^d),
 \label{2.8*}
\end{equation}
where the first arrow takes $(f,g)$ to the function $(f\otimes
g)(x_1,x_2)=f(x_1)g(x_2)$, the second arrow is the action of
operator~\eqref{2.6*}, and the third arrow is the restriction of
elements of $\mathscr{S}^\beta(\oR^{2d})$ to the diagonal
$x_1=x_2$. The first map is obviously continuous, and we now argue
that the third map is also continuous. Although the spaces
$\mathscr{S}^\beta$  are not invariant under the Fourier
transformation, they closely resemble the Schwartz space $\mS$ in
their other  properties. These spaces are complete and metrizable,
i.e., belong to the class of Fr\'echet spaces. Furthermore, they
are Montel spaces (or perfect in nomenclature of~\cite{GS2}) and
nuclear. An analogue of Schwartz's kernel theorem states that
$\mathscr{S}^\beta(\oR^{2d})$ coincides with the completion of the
tensor product $\mathscr{S}^\beta(\oR^d)\mathbin{\otimes_\pi}
\mathscr{S}^\beta(\oR^d)$ equipped with the projective topology.
Therefore, the set of continuous bilinear maps
$\mathscr{S}^\beta(\oR^d)\times \mathscr{S}^\beta(\oR^d)\to
\mathscr{S}^\beta(\oR^d)$  can be identified with the set of
continuous linear maps $\mathscr{S}^\beta(\oR^{2d})\to
\mathscr{S}^\beta(\oR^d)$. In particular, the linear map $\widehat
{\mathsf m}$ corresponds to the ordinary pointwise multiplication
$\mathsf m\colon (f,g)\to f\cdot g$, and its continuity  follows
from (and amounts to) the fact that $\mathscr{S}^\beta(\oR^d)$ is
a topological algebra under the ordinary multiplication. We thus
get the following theorem.

\medskip
 {\bf Theorem 3.} {\it The spaces  $\mathscr{S}^{\beta}(\oR^d)$
with $\beta\le1/2$ are topological algebras under the Moyal
$\star$-product. If $f,g \in \mathscr{S}^\beta(\oR^d)$, where
$\beta\le1/2$, then the series~\eqref{1.2*} is absolutely
convergent in this space.}

\medskip
Another way of proving this is  to estimate the expression
$(1+|x|)^N |\partial^\kappa(f\star g)(x)|$ with the use of
Leibniz's formula. Such a computation is almost identical to the
proof of theorem~4 in paper~\cite{S07} dealing with the spaces
$S^\beta_\alpha$.

 The Gelfand-Shilov spaces $S^\beta$ are constructible from the
 spaces $S_N^{\beta, B}$ in the following way:
\begin{equation}
S^\beta=\bigcup_{B>0} S^{\beta, B}, \qquad S^{\beta,
B}=\bigcap_{B'>B,N\in\oZ_+} S_N^{\beta, B'}.
 \label{2.9*}
\end{equation}
A sequence $f_n$ is said to be convergent to an element $f\in
S^\beta$ if there is a $B>0$ such that all $f_n$ and $f$ are
contained in the space $S^{\beta, B}$ and $f_n\to f$ in each of
its norms.

Gelfand and Shilov~\cite{GS2} have shown that the spaces $S^\beta$
are algebras under the pointwise multiplication and that this
operation is separately continuous  in their topology.
Mityagin~\cite{M} has proved that these spaces are nuclear.
Another proof is given in~\cite{Izv}, where in addition their
completeness  is established and the corresponding kernel theorem
is proved. From this theorem, it follows that the set of
separately continuous bilinear maps $S^\beta(\oR^d)\times
S^\beta(\oR^d)\to S^\beta(\oR^d)$ is identified with the set of
continuous linear maps $S^\beta(\oR^{2d})\to S^\beta(\oR^d)$. We
combine these facts in a manner analogous to that used in the case
of $\mS^\beta$ and suppose that $\beta$ satisfies the strict
inequality $\beta<1/2$. Then $e^{b|z|^{1/2}}\le C_\epsilon
e^{\epsilon|z|^{1/\beta}}$, where $\epsilon>0$ can be taken
arbitrarily small, and we obtain the following result.

\medskip
 {\bf Theorem 4.} {\it The operator~\eqref{2.6*} maps every
space $S^\beta(\oR^{2d})$ with $\beta<1/2$   to itself
continuously. The spaces $S^\beta(\oR^d)$, where $\beta<1/2$, are
algebras under the Moyal $\star$-product, and  the star
multiplication is separately continuous under their topology. If
$f,g \in S^\beta(\oR^d)$, then the series~\eqref{1.2*} converges
absolutely in this space.}

\medskip
The Fourier transformation $\mathcal{F}\colon f(x)\to\hat
f(p)=\int f(x)e^{ip\cdot x}dx$ converts $S^\beta$ to the space
$S_\beta$ which consists of all smooth functions satisfying the
inequalities
$$
|\partial^\kappa  h(p)|\le C_\kappa e^{-|p/B|^{1/\beta}}\qquad
\text{for some $B(h)>0$ and for every $\kappa$},
$$
whereas $\mathscr{S}_{\beta}=\mathcal{F}[\mathscr{S}^{\beta}]$
consists of the functions satisfying
$$
|\partial^\kappa h(p)|\le C_{\kappa, B}e^{-|p/B|^{1/\beta}}\quad
\text{for each $B>0$  and for every $\kappa$}.
$$
 The operator~\eqref{2.6*} turns into the multiplication of
 the Fourier transforms by the function
\begin{equation}
e^{-(i/2)p_1\theta p_2},\qquad \text{where}\quad p_1\theta
p_2\eqdef p_{1\mu}\theta^{\mu\nu}p_{2\nu}. \label{2.10*}
\end{equation}
Clearly, this function is a multiplier of
$\mathscr{S}_{\beta}(\oR^{2d})$ and of $S_\beta(\oR^{2d})$ for any
$\beta$. Hence the operator~\eqref{2.6*} admits  continuous
extension to all these spaces, and this extension is unique
because $S_0=C^\infty_0$ is dense  in each of them. It follows
that the $\star$-product also has a unique continuous extension to
the spaces with  $\beta>1/2$. This extension is defined by
 \begin{equation}
(f\times g)(x)=\frac{1}{(2\pi)^{2d}}\int\int \hat f(q)\hat
g(p)\,e^{-iqx-ipx-(i/2)q\theta p} dq dp \notag
\end{equation}
and is often called the ``twisted product''~\cite{G-BV}. The
uniqueness of the extension entitles us to identify the operation
$(f,g)\to f\times g$ with the $\star$-multiplication and to say
that $\mathscr{S}^{\beta}$ and $S^\beta$ are star product algebras
for any $\beta$. However, proposition~2 in~\cite{S07} shows that
every space $\mathscr{S}^{\beta}$ with $\beta>1/2$ and every
$S^\beta$ with $\beta\ge 1/2$ contain functions for which the
series~\eqref{1.2*} is not convergent in the topology of these
spaces.

\section{\large Test function algebras over cones in $\oR^d$}

The operator~\eqref{2.6*} is nonlocal, but when acting on the
functionals defined on  $\mS^\beta$ or on $S^\beta$, it preserves
the property of a rapid decrease along a given direction of
$\oR^{2d}$ if a functional has such a property.  In order for this
statement to be given a precise mathematical meaning, we use
spaces which are related to $\mS^\beta(\oR^d)$ and
$S^\beta(\oR^d)$, but associated  with cones in $\oR^d$.  Such
sheaves of spaces arise naturally in nonlocal quantum field
theory, see~\cite{FS92,S93}.

Let $U$ be an open connected cone in $\oR^d$. We denote by
$S^{\beta, B}_N(U)$  the space of all infinitely differentiable
functions on $U$ with the finite norm
\begin{equation}
\|f \|_{U,B,N}=\sup_{x\in
U;\kappa}\,(1+|x|)^N\frac{|\partial^\kappa
f(x)|}{B^{|\kappa|}\kappa^{\beta\kappa}}.
 \label{3.1*}
\end{equation}
The spaces $\mS^\beta(U)$, $S^{\beta, B}(U)$ and $S^\beta(U)$ are
constructed from $S^{\beta, B}_N(U)$ by formulae analogous
to~\eqref{2.7*} and \eqref{2.9*}. If $\beta\le 1$, then all
elements of these spaces can be continued analytically to the
whole of $\oC^d$ and this definition can be rewritten in terms of
complex variables. Using the Taylor and Cauchy formulae, it is
easy to verify (see, e.g., \cite{FS92} for details) that the space
$S^\beta(U)$ with $\beta<1$ coincides with the space of all entire
analytic functions satisfying the inequalities
  \begin{equation}  |f(z)|\le C_N (1+|x|)^{-N}\,e^{d(Bx,U)^{1/(1-\beta)}
  +|By|^{1/(1-\beta)}},\qquad N=0,1,\dots,
   \label{3.2*}
   \end{equation}
    where $z=x+iy$,
$d(x,U)=\inf_{\xi\in U}|x-\xi|$ is the distance from  $x$ to $U$
 and the constants $C_N, B$ depend on $f$. This space is independent of the
choice of the norm $|\cdot|$ on $\oR^d$, because all these norms
are equivalent.  We also note that $d(Bx,U)=Bd(x,U)$ since $U$ is
a cone. The analytic continuations of the elements of
$\mS^\beta(U)$ satisfy analogous inequalities for every $B$ and
for every $N$ with constants $C_{B,N}$ instead of $C_N$. This
representation makes it clear that the spaces $S^\beta(U)$ and
$\mS^\beta(U)$, where $\beta<1$, are algebras under the ordinary
multiplication.

The arguments used in the proofs of theorems 1 and 2 are
completely applicable to the spaces over cones and furnish the
following result.

\medskip
 {\bf Theorem 5.} {\it Let $U$ be an open cone in $\oR^{2d}$.
 If $B<1/\sqrt{e|\theta|}$, then the operator~\eqref{2.6*}
maps  the normed space  $S_N^{1/2, B}(U)$  to $S_N^{1/2,B'}(U)$,
where $B'= B\sqrt{2}$, and is  bounded. Each of the spaces
$\mathscr{S}^\beta(U)$ with $\beta\le1/2$ and $S^\beta(U)$ with
$\beta<1/2$ is continuously mapped by this operator into itself.
Consequently, it is defined and continuous on their dual spaces
$\mathscr{S}^{\prime\beta}(U)$, $S^{\prime\beta}(U)$.  The series
obtained by applying the operator~\eqref{2.6*} to the elements of
these spaces are absolutely convergent.}

\medskip
As is shown in~\cite{Izv},  the spaces  $S^{\beta, B}(U)$ are
nuclear. It immediately follows that  $\mS^\beta(U)$ and
$S^\beta(U)$ also have this  property. Theorem~6 of~\cite{Izv}
states that the space $S^\beta(U\times U)$ coincides with the
completion of the tensor product $S^\beta
(U)\mathbin{\otimes_\iota} S^\beta (U)$ endowed with the inductive
topology. Let   $f,g\in S^\beta(U)$. In complete analogy to the
reasoning of section~2, we can decompose the map $(f,g)\to f\star
g$ as follows:
\begin{equation}
S^\beta(U)\times
S^\beta(U)\stackrel{\otimes}{\longrightarrow}S^\beta(U\times U)
\stackrel{e^{(i/2)\partial_{x_1}\theta\partial_{x_2}}}
{\longrightarrow}S^\beta(U\times U)\stackrel{\widehat {\mathsf
m}}{\longrightarrow}S^\beta(U).
 \label{3.3*}
\end{equation}
The former map in~\eqref{3.3*} is  separately continuous and the
latter two are continuous. (As before, we denote by $\widehat
{\mathsf m}$ the linear map that canonically corresponds to the
ordinary product.) A similar representation certainly holds for
the spaces $\mS^\beta(U)$, and  in that case we have even a
simpler situation because these are Fr\'echet spaces and we need
not distinguish between  separately continuous and continuous
linear maps. We thus get the following theorem.

\medskip
{\bf Theorem 6.} {\it Let $U$ be an open cone in $\oR^d$. Every
space $\mathscr{S}^\beta(U)$ with $\beta\le1/2$ is a topological
algebra under the Moyal $\star$-product. If $\beta< 1/2$, then
$S^\beta(U)$ also is an algebra with respect to the Moyal product
and the  $\star$-multiplication is separately continuous under its
topology. The series~\eqref{1.2*} converges absolutely in these
spaces for each pair of their elements.}

\medskip
The special convenience of $S^\beta$ (and $S^\beta_\alpha$) is
that the generalized functions defined on these spaces of analytic
test functions have been shown to possess the property of angular
localizability, which is specified in the following manner. We say
that a functional $v\in S^{\prime\beta}(\oR^d)$ is carried by a
closed cone $K\subset \oR^d$ if $v$ admits a continuous extension
to every space $S^\beta(U)$, where $U\supset K\setminus\{0\}$.
This property is equivalent to the existence of a continuous
extension to the space
  \begin{equation}
S^\beta(K)=\bigcup_{U\supset K\setminus\{0\}}S^\beta(U)
    \label{3.4*}
   \end{equation}
endowed with the topology induced by the family of injections
$S^\beta(U)\to S^\beta(K)$. When such an extension exists, it is
unique because  $S^\beta$ is dense in $S^\beta(U)$ and in
$S^\beta(K)$ by theorem~5 of~\cite{S97}. The
representation~\eqref{3.2*} makes it clear that outside $K$
elements of $S^\beta(K)$ can have an exponential growth of order
$1/(1-\beta)$  and a finite type. Hence we are entitled to
interpret the existence of a nontrivial carrier cone of $v\in
S^\beta(\oR^d)$ as a falloff property of this functional in the
complementary cone or more specifically as a decrease  faster than
exponentially with order  $1/(1-\beta)$ and  maximum type.
Moreover, the  relation
 \begin{equation}
 S^{\prime\, \beta}(K_1\cap K_2)= S^{\prime\, \beta}(K_1)\cap S^{\prime\,
  \beta}(K_2)
    \label{3.5*}
   \end{equation}
holds, which implies that every element of $S^{\prime\,
\beta}(\oR^d)$ has a unique minimal closed carrier cone in
$\oR^d$. This fact has been established in~\cite{FS92,S93} for
$0<\beta<1$, and a detailed proof of the relation~\eqref{3.5*} for
the most complicated case $\beta=0$ is available in~\cite{Izv}.
Clearly, the spaces $S^\beta(K)$ over closed cones  also are
algebras under the Moyal $\star$-product if  $\beta<1/2$.

\section{\large Asymptotic commutativity and $\theta$-locality}

It is believed that a mathematically rigorous theory of quantum
fields on noncommutative spacetime shall adopt the basic
assumption of the axiomatic approach~\cite{SW,J,BLOT} that quantum
fields  are operator-valued generalized functions with a common,
dense and invariant domain $D$ in the Hilbert space of states. The
optimal test function spaces may be model-dependent, but the above
consideration shows that in any case the space $\mS^{1/2}$ and the
spaces $S^\beta$ with $\beta<1/2$, as well as their related spaces
over cones, are attractive for use in noncommutative QFT.

Analytic test functions were used in nonlocal field theory over
many years and it would be reasonable to draw this experience. The
 axiomatic formulation of nonlocal QFT developed in~\cite{FS92,
S93,S99} is based on the idea of changing  local commutativity
 to an asymptotic commutativity condition, which means that the
commutator or anticommutator  of any two fields of the theory is
carried by the cone
\begin{equation}
\overline{\oV}\times \oR^d =\{(x,x')\in \oR^{2d}\colon (x-x')^2\ge
0\}.
 \label{4.1*}
 \end{equation}
In more exact terms, if the fields    $\phi$, $\psi$ are defined
on the test function space $S^\beta(\oR^d)$, $\beta<1$, then
either

\begin{equation}
\langle\Phi\, [\phi(x),\psi(x')]_-\Psi\rangle
    \label{4.2*}
   \end{equation}
   or
\begin{equation}
\langle\Phi\, [\phi(x),\psi(x')]_+\Psi\rangle
    \label{4.3*}
   \end{equation}
is carried by the cone~\eqref{4.1*} for all $\Phi, \Psi\in D$. The
matrix elements~\eqref{4.2*}, \eqref{4.3*} can be regarded as
generalized functions on $\oR^d$ because  $S^\beta$ is nuclear and
the relation $S^\beta (\oR^d)\mathbin{\hat{\otimes}_\iota} S^\beta
(\oR^d) =S^\beta (\oR^{2d})$ holds. The asymptotic commutativity
condition becomes weaker with  decreasing $\beta$. For $\beta=0$,
it  means that the commutator of observable fields averaged with
test functions in $S^0$ decreases at spacelike separation no worse
than exponentially with order 1 and maximum type. Together with
other Wightman axioms, this condition ensures the existence of the
CPT-symmetry operator and the normal spin-statistics relation for
the nonlocal quantum fields.  The proofs~\cite{S99} of these
theorems use the notion of the analytic wave front set of
distributions  in an essential way. This generalization of the
local commutativity axiom also preserves the cluster decomposition
property of the vacuum expectation values. As  shown
in~\cite{S82}, it preserves even the strong exponential version of
this property if the theory has a mass gap. This makes possible
interpreting the nonlocal QFT subject to the asymptotic
commutativity condition in terms of the particle scattering
because the cluster decomposition property plays a key role in
constructing the asymptotic states and the S-matrix.

In~\cite{S06}, we  discussed  some peculiarities of using the
analytic test functions in quantum field theory on noncommutative
spacetime for the case of a charged scalar field and space-space
noncommutativity. We have shown that this theory has CPT-symmetry
if it satisfies a suitably modified condition of asymptotic
commutativity. This modification  uses the
generalization~\cite{Sm, Izv} of the notion of carrier cone to the
bilinear forms.

The test function spaces $S^\beta$, $\beta<1/2$, are convenient
for use in quantum field theory on noncommutative spacetime
because they are algebras under the  $\star$-product and the
generalized functions defined on them have the property of angular
localizability, which enables one to apply analogues of some basic
theorems of Schwartz's theory of distributions. Moreover,
$S^\beta(\oR^d)$ are invariant under the affine transformations of
coordinates and the spaces of this kind over the light cone are
invariant under the Poincar\'e group. The asymptotic commutativity
provides a way of formulating causality in noncommutative QFT, but
it is insensitive to the magnitude of the noncommutativity
parameter which determines the fundamental length scale. The above
analysis suggests that a more accurate formulation can be obtained
by using spaces $S^{1/2,B}$. The nonlocal effects in quantum field
theory on noncommutative spacetime are  determined by the
structure of the Moyal $\star$-product, and one might expect that
in this theory each of the matrix elements~\eqref{4.2*}
(or~\eqref{4.3*} for unobservable fields) admits a continuous
extension to the space
\begin{equation}
S^{1/2,B}(\overline{\oV}\times \oR^d),\qquad \text{where}\quad
B\sim \frac{1}{\sqrt{|\theta|}}\,.
    \label{4.4*}
   \end{equation}
(In general, $B$ may  depend on the fields $\phi$, $\psi$ and the
states $\Phi$, $\Psi$.) This condition will be called
$\theta$-{\it locality}. Clearly, it is stronger than the
asymptotic commutativity condition stated for $\beta<1/2$, but it
is also consistent with the Poincar\'e covariance. Conceivably,
the $\theta$-locality expresses the absence of acausal effects on
scales much larger than the fundamental scale $\Lambda\sim
\sqrt{|\theta|}$. If such is the case,  this assumption might be
called macrocausality. It should be emphasized that we do not
assume here that the fields are defined only on the analytic test
functions. It is quite possible that their matrix elements are
usual tempered distributions. In other words, we use $S^{1/2,B}$
as a tool for formulating causality rather than as the functional
domain of definition of fields. In the next section, we reconsider
from this standpoint a typical example which was used
in~\cite{Ch,G} for showing the violation of microcausality in
quantum field theory on noncommutative spacetime.

\section{\large An example}

Let $\phi$ be a free neutral scalar field of mass $m$ in a
spacetime of $d$ dimensions and let
\begin{multline}
\mathcal O(x)\equiv:\phi\star\phi:(x) =\lim_{x_1,x_2\to
x}:\phi(x_1)\phi(x_2):\\+\sum_{n=1}^\infty\left(\frac{i}{2}
\right)^n\frac{1}{n!}\,\theta^{\mu_1\nu_1}\dots
\theta^{\mu_n\nu_n}\lim_{x_1,x_2\to x}:\partial_{\mu_1}\dots
\partial_{\mu_n}\phi(x_1)\,\partial_{\nu_1}\dots\partial_{\nu_n}\phi(x_2):.
 \label{5.1*}
\end{multline}
 Every term in~\eqref{5.1*} is well defined as a Wick binomial.
M.~Chaichian {\it et al}~\cite{Ch} and O.~Greenberg~\cite{G}
studied the question of microcausality in noncommutative QFT for
the choice $\mathcal O$ as a sample observable. Specifically, they
considered the matrix element
$$
\langle 0|\,[\mathcal O(x), \mathcal O(y)]_-|p_1,p_2\rangle
$$
at $x^0=y^0$. In the case of space-space noncommutativity, with
$\theta^{12}=-\theta^{21}\ne 0$ and the other elements of the
$\theta$-matrix equal to zero, the commutator $[\mathcal O(x),
\mathcal O(y)]_-$ vanishes in the light wedge
$(x^0-y^0)^2<(x^3-y^3)^2$, but Greenberg found that $[\mathcal
O(x), \partial_\nu\mathcal O(y)]_-$ fails to vanish outside this
wedge and so violates microcausality. We shall show that
nevertheless the $\theta$-locality condition is   fulfilled for
this observable and this result holds irrespectively  of the type
of noncommutativity.

First, we calculate the vacuum expectation value
\begin{equation}
\mW(x,y;z_1,z_2)=\langle 0|\,\mathcal O(x) \mathcal O(y)
:\phi(z_1)\phi(z_2):|0\rangle.
 \label{5.3*}
 \end{equation}
We use the Wick theorem and express
$$
\langle
0|:\phi(x_1)\phi(x_2)\!:\,\,:\phi(y_1)\phi(y_2)\!:\,\,:\phi(z_1)\phi(z_2)\!:|0\rangle
$$
in terms of the two-point function
\begin{equation}
w(x-y)= \langle
0|\phi(x)\phi(y)|0\rangle=\frac{1}{(2\pi)^{d-1}}\int e^{-ik\cdot
(x-y)}\vartheta(k^0)\delta(k^2-m^2)\,dk.
 \notag
\end{equation}
Applying then the relation
\begin{equation}
\lim_{x_1, x_2 \rightarrow
x}\exp\left(\frac{i}{2}\partial_{x_1}\theta\,\partial_{x_2}\right)
e^{ik\cdot x_1}e^{ip\cdot x_2}
 \equiv e^{ik\cdot x}\star e^{ip\cdot x}=e^{-(i/2)k\theta p}e^{i(k+p)\cdot
 x},
 \notag
\end{equation}
 we obtain
\begin{multline}
\mW(x,y;z_1,z_2)= 4\!\int\!\!
\frac{dkdp_1dp_2}{(2\pi)^{3(d-1)}}\,\vartheta(k^0)\delta(k^2-m^2)
\prod_{i=1}^2\vartheta(p_i^0)\delta(p_i^2-m^2)
\cos\left(\frac{1}{2}k\theta   p_i\right)\\\times e^{-ik\cdot
(x-y)-ip_1\cdot (x-z_1)-ip_2\cdot (y-z_2)} + (z_1\leftrightarrow
z_2).
 \label{5.4*}
\end{multline}
This formal derivation should be  accompanied by a comment. The
function
$$\cos\left(\frac{1}{2}k\theta
p_1\right)\cos\left(\frac{1}{2}k\theta p_2\right)
$$
 is a multiplier
for the Schwartz space $\mS$, and hence the right-hand side
of~\eqref{5.4*} is well defined as a tempered distribution. This
distribution is obtained by applying the operator
\begin{equation}
\cos\left(\frac{1}{2}\partial_x\theta
   \partial_{z_1}\right)\cos\left(\frac{1}{2}\partial_y\theta
   \partial_{z_2}\right)
 \label{5.6*}
\end{equation}
to the distribution
$$
4\!\int\!\!
\frac{dkdp_1dp_2}{(2\pi)^{3(d-1)}}\,\vartheta(k^0)\delta(k^2-m^2)
   \prod_{i=1}^2 \vartheta(p_i^0)\delta(p_i^2-m^2)
e^{-ik\cdot (x-y)-ip_1\cdot (x-z_1)-ip_2\cdot (y-z_2)}\\ +
(z_1\leftrightarrow z_2).
$$
By theorem~2, the operator~\eqref{5.6*} is defined and continuous
on $\mS^{1/2}(\oR^{4d})$ (and on any space $S^\beta(\oR^{4d})$
with $\beta<1/2$) and the power series expansion of~\eqref{5.4*}
in $\theta$ is weakly convergent to $\mW$ in the dual space
$\mS^{\prime 1/2}$. This implies the strong convergence because
$\mS^{1/2}$ is a Montel space. However, that is not to say that
this expansion converges to $\mW$  in the topology of the space
$\mS'$ of tempered distributions.

Using~\eqref{5.4*}, we obtain
\begin{multline}
\langle 0|\,[\mathcal O(x),\mathcal
O(y)]_-:\phi(z_1)\phi(z_2)\!:|0\rangle\\= 4\!\int\!\!
\frac{dkdp_1dp_2}{(2\pi)^{3(d-1)}}\,\epsilon (k^0)\delta(k^2-m^2)
\prod_{i=1}^2\theta(p_i^0)\delta(p_i^2-m^2)
\cos\left(\frac{1}{2}k\theta  p_i\right)\\\times e^{-ik\cdot
(x-y)-ip_1\cdot (x-z_1)-ip_2\cdot (y-z_2)} + (z_1\leftrightarrow
z_2),
 \label{5.7*}
\end{multline}
which agrees with  formula (7) of~\cite{G}.

\medskip
{\bf Theorem 7.} {\it The restriction of the
distribution~\eqref{5.7*} to $S^{1/2}$ has a continuous extension
to the space $S^{1/2, B}(\oV\times \oR^{3d})$, where
$B<1/\sqrt{e|\theta|}$ and
\begin{equation}
\oV\times \oR^{3d}=\{(x,y,z_1,z_2)\in \oR^{4d}\colon (x-y)^2>0\}.
 \label{5.8*}
\end{equation}
A fortiori, the restriction of this distribution to any space
$S^\beta(\oR^{4d})$ with $\beta<1/2$ is strongly carried by the
closed cone $\overline{\oV}\times \oR^{3d}$.

\medskip
Proof.} Let $B'= B\sqrt{2}$. The restriction of~\eqref{5.7*} to
$S^{1/2,B'}(\oR^{4d})$ is obtained by applying the
operator~\eqref{5.6*} to the restriction of
\begin{multline}
D(x,y;z_1,z_2)\equiv \langle
0|\,[:\phi^2\!:(x),:\phi^2\!:(y)]_-:\phi(z_1)\phi(z_2):|0\rangle\\=
 4i\Delta(x-y)w(x-z_1)w(y-z_2)+ (z_1\leftrightarrow z_2).
 \label{5.9*}
\end{multline}
Clearly,  $D(x,y;z_1,z_2)$ vanishes for $(x-y)^2<0$ and the
restriction $D| S^{1/2,B'}(\oR^{4d})$ has a continuous extension
$\widetilde{D}$ to  the space $S^{1/2,B'}(\oV\times \oR^{3d})$.
This extension can be defined by $(\widetilde{D},f)=(D,\chi f)$,
where $\chi$ is a multiplier of the Schwartz space, which is equal
to 1  on an $\epsilon$-neighborhood of $\bar\oV\times \oR^{3d}$
and  to zero outside the
 $2\epsilon$-neighborhood. Such a multiplier satisfies the uniform
 estimate
 $|\partial^\kappa\chi|\le C_\kappa$, and the multiplication by $\chi$
  maps
$S^{1/2,B'}(\oV\times \oR^{3d})$ into $\mS(\oR^{4d})$
continuously. By theorem~5, applying~\eqref{5.6*} to
$\widetilde{D}$, we obtain a continuous extension of the
functional~\eqref{5.7*} to the space $S^{1/2, B}(\oV\times
\oR^{3d})$. This proves theorem~7. We point out once again that
this theorem holds for any matrix $\theta$  and in particular for
both space-space and time-space noncommutativity.

\section{\large Concluding remarks}

Our analysis shows that the $\theta$-locality condition or the
weaker condition of asymptotic commutativity for the restrictions
of fields to the test function spaces $S^\beta$, $\beta<1/2$, can
serve as  a substitute of microcausality in quantum field theory
on noncommutative spacetime even though the fields are tempered.
The character of singularity is certainly dependent on the model,
but multiplication by the exponential~\eqref{2.10*} alone cannot
spoil  temperedness. As  stressed in~\cite{FW,S}, any attempt to
replace microcausality by a weaker requirement must take  the
theorem on the global nature of local commutativity into
consideration. The Borchers and Pohlmeyer version~\cite{BP} of
this theorem states that local commutativity follows from an
apparently weaker assumption that $[\phi(x),\psi(x')]_\pm$
decreases at large spacelike separation faster than exponentially
of order 1. The example $:\phi\star\phi:$ discussed above
demonstrates that this theorem is inapplicable to the asymptotic
commutativity condition  and that this condition does not imply
local commutativity. The point is that the fast decrease at
spacelike separation is understood here differently than
in~\cite{BP}, as a property of the field (anti)commutators
averaged with appropriate test functions. We have restricted our
consideration to the specific matrix element of the commutator,
but the technique developed in~\cite{SS01} enables one to
construct the operator realization of $:\phi\star\phi:$ in the
state space of $\phi$ and to prove that in this instance the
$\theta$-locality condition is completely fulfilled. In
combination with the usual relativistic transformation law of
states and fields, the asymptotic commutativity ensures  the
existence of CPT-symmetry and the normal spin-statistics relation
for nonlocal fields~\cite{S99}. One might expect that in
noncommutative QFT similar conclusions can be deduced  from a
suitable combination of the $\theta$-locality and the twisted
Poincar\'e covariance~\cite{FW,CK} which is currently received
much attention.

Most, if not all, of the results established above for $S^\beta$
can readily be extended to the spaces $S^\beta_\alpha$ whose
topological structure is even simpler.  In particular, a theorem
similar to theorem~1 holds with $S^{1/2, B}_{\alpha, A}$ in place
of $S_N^{1/2, B}$. Analogues of theorems~2 and~3 hold for
$\mathscr{S}^{\beta}_\alpha=\bigcap_{A,B}S^{\beta,B}_{\alpha,A}$,
where $\beta\le1/2$ and $\alpha>1-\beta$. An analogue of theorem~4
is valid for
$S^{\beta}_\alpha=\bigcup_{A,B}S^{\beta,B}_{\alpha,A}$ with
$\beta<1/2$ and $\alpha\ge 1- \beta$. Of course, analogues of
theorems~5 and 6 hold with the same replacements.

\medskip
\section* {\large Acknowledgments}

 This paper was supported in part by the the Russian
Foundation for Basic Research (Grant No.~05-01-01049) and the
Program for Supporting Leading Scientific Schools (Grant
No.~LSS-4401.2006.2).

\end{document}